\documentclass[smallabstract,smallcaptions]{dccpaper}

\usepackage{amsmath}
\usepackage{amssymb}
\usepackage{amsxtra}
\usepackage{latexsym}
\usepackage{cite}
\usepackage{booktabs}
\usepackage[pdfpagemode=UseNone,pdfstartview=FitH,pdfview=FitH,colorlinks=true,
 pdftitle=Daala:\ A\ Perceptually-Driven\ Next\ Generation\ Video\ Codec,
 pdfauthor=Xiph.Org]{hyperref}
\usepackage{color}
\usepackage{url}
\usepackage{caption}
\usepackage{subfigure}
\usepackage{graphicx}

\newlength{\figurewidth}
\newlength{\smallfigurewidth}

\begin{document}

\title
{\large
\textbf{Daala: A Perceptually-Driven Next Generation Video Codec}
}

\author{%
Thomas J. Daede$^{\ast\dag}$, Nathan E. Egge$^{\ast\dag}$,
 Jean-Marc Valin$^{\ast\dag}$, Guillaume Martres$^{\ast\ddag}$, \\
 Timothy B. Terriberry$^{\ast\dag}$\\[0.5em]
{\small\begin{minipage}{\linewidth}\begin{center}
\begin{tabular}{ccccc}
$^{\ast}$Xiph.Org Foundation &\hspace*{0.25in} & $^{\dag}$Mozilla & \hspace*{0.25in} & $^{\ddag}$EPFL \\
21 College Hill Road && 331 E. Evelyn Ave. && Route Cantonale \\
Somerville, MA 02144 && Mountain View, CA 94041 && 1015 Lausanne \\
United States && United States && Switzerland \\
\multicolumn{5}{c}{\url{tterribe@xiph.org}} \\
\\
\multicolumn{5}{c}{Copyright 2015-2016 Mozilla Foundation.
This work is licensed under
 \href{https://creativecommons.org/licenses/by/4.0/}{CC-BY 4.0}.}
\\
\end{tabular}
\end{center}\end{minipage}}
}

\maketitle
\thispagestyle{empty}

\begin{abstract}
The Daala project is a royalty-free video codec that attempts to compete with
 the best patent-encumbered codecs.
Part of our strategy is to replace core tools of traditional video codecs with
 alternative approaches, many of them designed to take perceptual aspects into
 account, rather than optimizing for simple metrics like PSNR.
This paper documents some of our experiences with these tools, which ones
 worked and which did not, and what we've learned from them.
The result is a codec which compares favorably with HEVC on still images, and
 is on a path to do so for video as well.
\end{abstract}

\Section{Introduction}

Since its inception, Daala has used lapped transforms~\cite{TDLT}.
These promise to structurally eliminate blocking artifacts from the transform
 stage, one of the most annoying artifacts at low bitrates~\cite{MS89,Tra01}.
Although extended to variable block sizes and block sizes up to $16\times 16$
 for still images~\cite{DLT05}, extensions to the larger block sizes found in
 modern codecs become problematic due to the exponential search complexity.
We now use a fixed-lapping scheme that admits an easy search and improves
 visual quality despite a lower theoretic coding gain.

Daala also employs OBMC~\cite{OBMC} to eliminate blocking artifacts from the
 prediction stage.
Early work demonstrated improvements simply by running OBMC as a post-process
 to simple block-matching algorithms~\cite{WS91}.
This was later extended to multiple block sizes, including adaptively
 determining the overlap~\cite{ZAS98}, but still running as a post-process.
We use a novel structure borrowed from surface simplification literature that
 allows efficient searching and partition size selection using the actual
 prediction, instead of a block-copy approximation.

Daala also builds on the vector quantization work of the Opus audio
 codec~\cite{Opus}, extending its gain-shape quantization scheme to support
 predictors and adaptive entropy coding~\cite{PVQ}.
We explicitly encode the gain of bands of AC coefficients, that is, their
 contrast, and explicitly code how well each band matches the predictor.
By extracting a small number of perceptually meaningful parameters like this
 from an otherwise undifferentiated set of transform coefficients, this
 ``Perceptual Vector Quantization'' enables a host of new techniques.
We have demonstrated its use for automatic activity masking (to preserve detail
 in low-contrast regions)~\cite{PVQ} and frequency-domain Chroma-from-Luma
 (CfL) to enhance object boundaries in the color planes~\cite{CfL}.

Both lapped transforms and PVQ are particularly susceptible to ringing
 artifacts, the former due to the longer basis functions and the latter due to
 the tendency either to skip entire diagonal bands (giving artifacts on
 diagonal edges similar to wavelets) or to inject energy into the wrong place
 when trying to preserve contrast.
Furthermore, lapping prevents strong directional intra prediction, which could
 be used to create clean edges.
Therefore, we designed a sophisticated directional deringing filter, which
 aggressively filters directional features with minimal side information.

\Section{Methodology}

This section attempts to describe why we made many of the choices we did.
All of the code, including the full commit history, is available in our public
 git repository~\cite{daala-git}.
Where appropriate, it includes the four metrics we commonly examine, PSNR,
 SSIM~\cite{WVSS04}, PSNR-HVS-M~\cite{PSECAL07}, and multiscale
 FastSSIM~\cite{CB11}.
Unless otherwise specified, Bj\o ntegaard-delta~\cite{draft-testing} (BD) rate
 changes and other results are from our automated testing
 framework~\cite{AWCY}.
By default this uses 18 sequences ranging in resolution from $416\times 240$ to
 $1920\times 1080$ and $48$~to $60$~frames in length.

Overall, the codec constructs a frequency-domain predictor for each block,
 codes the input with PVQ using this predictor, and then filters the result.
In intra frames, we construct the predictor with simple horizontal and
 vertical prediction in the luma plane (copying coefficients) and
 Chroma-from-Luma in the chroma planes (described below).
In inter frames, we construct a motion-compensated reference frame for the
 whole image using OBMC, and then apply our forward transform to obtain the
 required frequency-domain predictor.
Discussion of the above techniques follows.

\SubSection{Lapped Transforms}

Video codecs have used adaptive filters to remove blocking artifacts since the
 H.263 standard was developed, at least.
However, there are also non-adaptive solutions to the blocking problem: lapped
 transforms.
Daala uses the time-domain construction from~\cite{Tra01}, with the DCT and
 lapping implemented using reversible lifting transforms.

Originally, we applied lapping in an order similar to that of a loop filter,
 applying the post-filter to rows of pixels first, for the entire image, and
 then columns (the pre-filter ran in the opposite order).
This allows maximal parallelism with minimal buffering.
However, this has two issues.
First, it creates strangely-shaped basis functions in the presence of varying
 block sizes.
Although possible to see in synthetic examples, we never observed an issue
 on real images, and adaptive deblocking filters have a similar issue.
Second, it makes block size decisions NP-hard, because you must know the sizes
 of both the current block and its neighbors in order to determine the amount
 of lapping to apply, creating a two-dimensional dependency graph that does not
 admit a tree structure, and thus has no dynamic programming solution.

We developed a heuristic to make up-front block size decisions, without a
 rate-distortion optimization (RDO) search, based on the estimated visibility
 of ringing artifacts.
However, it made clearly sub-optimal decisions for video, often choosing large
 transforms when only a small portion of a block changed.
To make a real RDO search tractable, we made two adjustments.
First, we made the order recursive: we first apply the lapping to the exterior
 edges of a $32\times 32$ superblock, then if we are splitting to a
 $16\times 16$, we filter the interior edges.
This is essentially the same as the order proposed by Dai et al.~\cite{DLT05}.
Second, we fixed the lapping size to an 8-point filter (4 pixels on either side
 of a block edge).
For $4\times 4$ blocks, we apply 4-point lapping to the interior edges of an
 $8\times 8$ block (overlapping with the 8-point lapping applied to the
 exterior edges).
When subsampled, the chroma planes always use 4-point lapping.

This removed the dependency on the neighbors' block sizes, allowing for a
 tree-structured dynamic programming search.
This proceeds bottom up.
At each level, we start with the optimal block size decision using blocks of
 at most $N/2\times N/2$, undo the lapping on the interior edges of the
 quadtree split, and compare with a single $N\times N$ block.
In both cases the exterior edges of the $N\times N$ block remain lapped, but we
 can at least make an apples-to-apples comparison of the relative distortions.

The result is an optimal solution using blocks of at most $N\times N$.
This optimization procedure produced BD-rate reductions of 10.4\% for PSNR, and
 12.3\% for SSIM.
On our more perceptual metrics, the changes were smaller: 4.5\% for PSNR-HVS-M
 and 5.2\% for multiscale FastSSIM.
This reflects the reduction in coding gain from the reduced lapping sizes.
These gains are almost entirely due to the improved decisions.
Using the same decisions produced by this fixed-lapping process with the
 previous variable-lapping scheme gave almost as much gain.

Despite the smaller lapping size, this scheme actually increases ringing
 artifacts.
We primarily code edges with $4\times 4$ blocks, but have increased the support
 of the $4\times 4$ blocks from 8~pixels to 12.
To reduce the effects of ringing, we are currently using only 4-point
 lapping on all block edges.
Making this change provided another 3.6\% gain in PSNR, and 1.4\% on
 PSNR-HVS-M, but lost 0.4\% on multiscale FastSSIM.
The visual impact is somewhat larger, and seems to be a regression for intra
 frames, but a win in some cases on video.
This will require more systematic visual testing.

%

\SubSection{Bilinear Smoothing Filter}

Because the smaller lapping size no longer completely eliminates blocking,
 especially in smooth gradients, we apply a bilinear smoothing filter to all
 $32\times 32$ blocks in keyframes.
The filter simply computes a bilinear interpolation between the four corners of
 a decoded block (after unlapping).
Then, it blends the result of that interpolation with the decoded pixels.
Unlike a conventional deblocking filter, it does not look outside of the
 current block at all.

For a given quantization step size, $Q$, the optimal Wiener filter gain is
\begin{align}
w & = \min\left(1, \frac{\alpha Q^2}{12D^2}\right)\ ,
\end{align}
 where $\alpha$ is a strength parameter (currently set to $5$ for luma and $20$
 for chroma) and $D^2$ is the mean squared error between the decoded block and
 the bilinear interpolation.
However, in practice we found that using $w^2$ works better than $w$, as
 it applies less smoothing when we are uncertain if it is a good idea.
The result is actually a small ($<$ 1\%) regression on metrics on
 \texttt{subset3}~\cite{subset3}, our large still image training subset, but
 provides a substantial visual reduction in blocking in gradients at low rates.

\begin{figure}[t]
\begin{center}
\subfigure[No bilinear smoothing.]{
 \label{fig:bilinear-off}
 \includegraphics[width=1.75in]{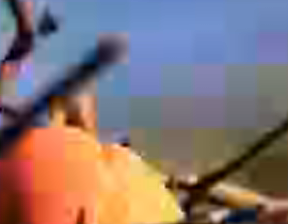}
}
\subfigure[With bilinear smoothing.]{
 \label{fig:bilinear-on}
 \includegraphics[width=1.75in]{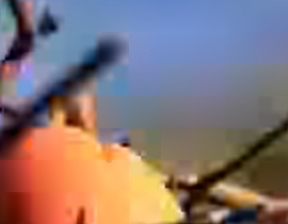}
}
\end{center}
\caption{\label{fig:bilinear-example}%
Excerpt from a 1.1~MP image compressed to 4.8~kB (0.035~bpp)
 \ref{sub@fig:bilinear-off} without the bilinear smoothing filter and
 \ref{sub@fig:bilinear-on} with the bilinear smoothing filter.
}
\end{figure}

\SubSection{Perceptual Vector Quantization}

Perceptual Vector Quantization uses the gain-shape vector
 quantization found in Opus~\cite{Opus}, extended to take advantage of
 prediction and adaptive entropy coding.
The main idea is that it splits the AC coefficients into bands and explicitly
 codes the magnitude (``gain''), $g$, of each band.
Then, the band is normalized to a unit vector, and the ``shape'' of the
 spectrum is encoded separately.
To handle prediction, we apply a Householder reflection to map the normalized
 prediction onto one of the axes, and then code the angle between that axis
 and the vector being quantized, $\theta$.
What remains is a vector on a sphere of dimension $N-1$ (where $N$ here is
 the size of the band) with a known radius $g\sin\theta$.
We normalize and code this vector using Pyramid Vector
 Quantization~\cite{Fis86} to reduce the number of degrees of freedom to
 $N-2$, eliminating any redundancy with $g$ and $\theta$.
For details, we refer readers to~\cite{PVQ}.

One thing this provides is explicit energy preservation, which is less
 important for video than audio, and can even be relaxed to save bits during
 RDO.
More importantly, the gain directly encodes the contrast in each band.
This allows us to implement activity masking without sending any extra side
 information.
Instead of using a linear quantizer, we compand the gain to get more resolution
 for smaller gains.
Then, once we know the gain, we adjust the quantizer used for $\theta$ and the
 shape vector to give more bits to smooth regions where errors are easily
 visible, and fewer bits to textured regions where they are not.
We disable activity masking on $4 \times 4$ blocks to avoid over-penalizing
 edges.

\SubSection{Chroma-from-Luma Prediction}

Although color conversion to $\mathrm{Y'C_bC_r}$ decorrelates the color channels globally
 across the frame, there is still some local correlation.
Lee and Cho created a spatial domain chroma predictor using a linear model of
 the relationship between chroma and luma built from previously-decoded
 neighbors~\cite{LeeCho09}.
Because Daala uses lapped transforms, these neighbors are unavailable when they
 are needed.
Although it would be possible, with additional complexity, to use more distant
 neighbors, this technique is even more efficient in the frequency domain.
PVQ even makes it possible to remove the model fitting step
 entirely.  For more details, see~\cite{CfL}.

We predict the chroma shape in PVQ directly from the luma coefficients.
We assume that frequency-domain chroma AC coefficients $C_{AC}(u,v)$ are
 linearly related to their co-located luma AC coefficients $L_{AC}(u,v)$, and
 that both have zero mean.
This linear model is exactly what is needed to create a chroma {\em shape}
 predictor, ${\bf r}$, for a band of AC coefficients from the
 reconstructed luma coefficients in that band, $\mathbf{\hat{L}}_{AC}$.
\begin{align}
C_{AC}(u,v) & = \alpha_{AC}\cdot L_{AC}(u,v)
 \implies {\bf r} = \alpha_{AC}\cdot\mathbf{\hat{L}}_{AC}
\end{align}
While the value of $\alpha_{AC}$ can be learned in the decoder, it is sometimes
 wrong.
It is cheaper to simply code its sign (its magnitude is already represented by
 PVQ's $g\sin \theta$).
This technique allows us to predict features within a block that cannot be
 predicted by straight edge extension.
Although this significantly reduces the number of bits spent on chroma, the
 perceptual impact is even larger, giving cleaner edges than the horizontal
 and vertical intra prediction we use for the luma plane.

\begin{figure}[t]
\begin{center}
\subfigure[Original chroma.]{
 \label{fig:chroma-orig}
 \includegraphics[width=1.75in]{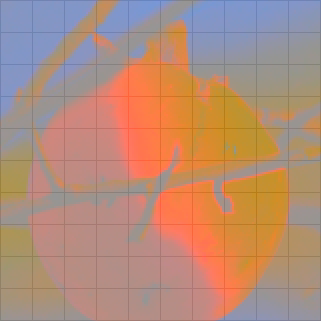}
}
\subfigure[With H/V prediction.]{
 \label{fig:chroma-hv}
 \includegraphics[width=1.75in]{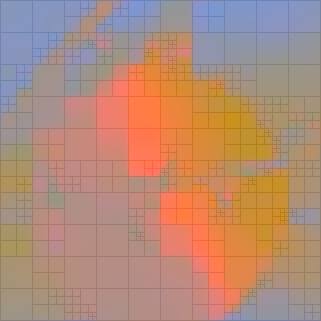}
}
\subfigure[With CfL.]{
 \label{fig:chroma-cfl}
 \includegraphics[width=1.75in]{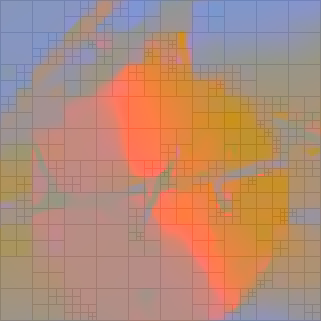}
}
\end{center}
\caption{\label{fig:pvq-cfl}%
Example comparison of \ref{sub@fig:chroma-orig} the original chroma planes
 with \ref{sub@fig:chroma-hv} horizontal and vertical spatial prediction and
 \ref{sub@fig:chroma-cfl} Chroma-from-Luma prediction.
}
\end{figure}

\SubSection{Deringing Filter}

The use of lapped transforms and the lack of directional intra prediction,
 along with the energy preservation of PVQ~\cite{PVQ}, make Daala particularly
 susceptible to ringing artifacts.
To combat this, Daala uses an in-loop deringing filter that takes into account
 the direction of edges and patterns being filtered.
On keyframes, this runs before the bilinear smoothing filter.
The filter identifies the direction of each block and adaptively filters along
 that direction.
A second filter runs across the lines filtered by the first filter, with more
 conservative thresholds to avoid blurring edges.
We describe the process in some detail here, since it has not been published
 elsewhere.

First, the decoder splits the image into $8\times 8$ blocks, and determines a
 dominant direction for each block from the decoded image.
These directions do not need to be transmitted, reducing the overhead of side
 information for the filter.
A perfectly directional block would have a constant value along all lines in a
 given direction.
The decoder minimizes the ``mean squared difference'' (MSD) between the decoded
 block and a perfectly directional block formed by taking the mean of the
 pixels in each line.

\begin{figure}[t]
\begin{center}
\subfigure[Example line partitions.]{
 \label{fig:Lines-for-direction}
 \raisebox{-.475\height}{\includegraphics[width=2in]{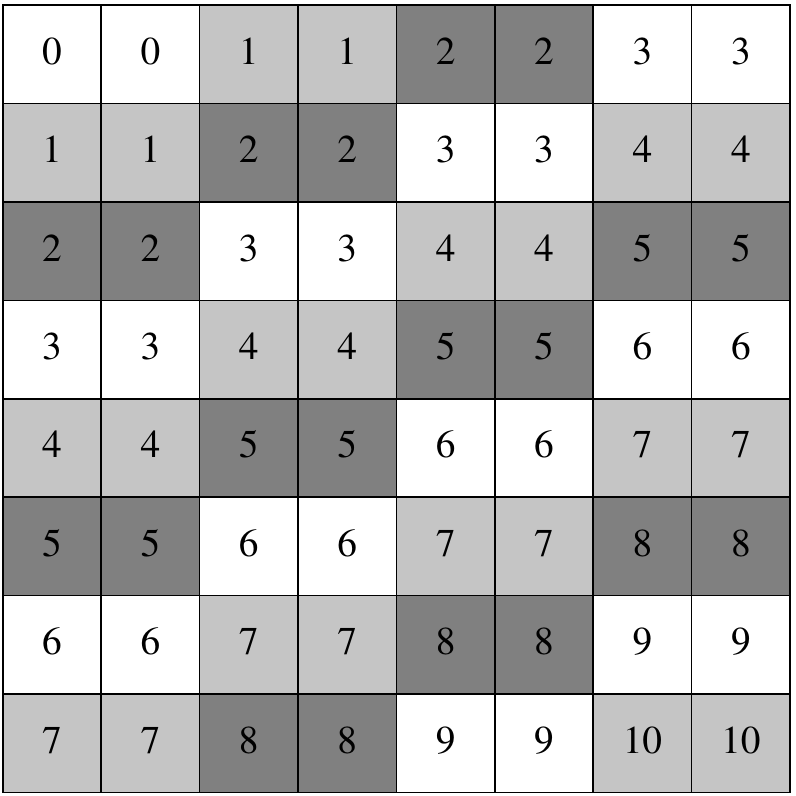}}
}
\subfigure[Direction parameters.]{%
  \label{tab:Direction-parameters}%
  \begin{tabular}{ccrr}
    \toprule
    Index & Direction & $d_x$ & $d_y$ \\\midrule
    0 & \rotatebox[origin=c]{45}{$\rightarrow$} &            1   &           -1   \\
    1 & \rotatebox[origin=c]{22.5}{$\rightarrow$} &            1   & $-\frac{1}{2}$ \\
    2 & \rotatebox[origin=c]{0}{$\rightarrow$} &            1   &            0   \\
    3 & \rotatebox[origin=c]{-22.5}{$\rightarrow$} &            1   &  $\frac{1}{2}$ \\
    4 & \rotatebox[origin=c]{-45}{$\rightarrow$} &            1   &            1   \\
    5 & \rotatebox[origin=c]{-67.5}{$\rightarrow$} &  $\frac{1}{2}$ &            1   \\
    6 & \rotatebox[origin=c]{-90}{$\rightarrow$} &            0   &            1   \\
    7 & \rotatebox[origin=c]{-112.5}{$\rightarrow$} & $-\frac{1}{2}$ &            1   \\
    \bottomrule
  \end{tabular}
}
\subfigure[Second stage.]{%
  \label{tab:Ortho-parameters}%
  \begin{tabular}{crr}
    \toprule
    Direction & $d_x$ & $d_y$ \\\midrule
    \rotatebox[origin=c]{-90}{$\rightarrow$} & 0 & 1\\
    \rotatebox[origin=c]{-90}{$\rightarrow$} & 0 & 1\\
    \rotatebox[origin=c]{-90}{$\rightarrow$} & 0 & 1\\
    \rotatebox[origin=c]{-90}{$\rightarrow$} & 0 & 1\\
    \rotatebox[origin=c]{-90}{$\rightarrow$} & 0 & 1\\
    \rotatebox[origin=c]{0}{$\rightarrow$} & 1 & 0\\
    \rotatebox[origin=c]{0}{$\rightarrow$} & 1 & 0\\
    \rotatebox[origin=c]{0}{$\rightarrow$} & 1 & 0\\
    \bottomrule
  \end{tabular}
}
\end{center}
\caption{
Line numbers for pixels following one direction in an $8\times 8$ block and the
 steps between pixels for each direction for each filter stage.
Pixels are always sampled using nearest-neighbor filtering, with no subpel
 filter.
}
\end{figure}

For each direction, $d$, we partition the pixels into distinct lines, as
 illustrated in Fig.~\ref{fig:Lines-for-direction}, indexing the lines by $k$.
The MSD, $\sigma_d^2$, is then
\begin{align}
 \sigma_d^2 & = \frac{1}{N}\sum_{k = 0}^{N_d - 1}
 \left[ \sum_{p \in P_{d,k}} \left(x_p - \mu_{d,k}\right)^2\right]
 \ ,\label{eq:direction-variance0}
\end{align}
 where $P_{d,k}$ is the set of pixels in line $k$ following direction $d$,
 $x_p$ is the value of the pixel at location $p$, $N_d$ is the number of lines
 in the block with direction $d$, and $N$ is the size of the block.
$\mu_{d,k}$ is the pixel average for the $k^\mathrm{th}$ line in direction $d$:
\begin{align}
\mu_{d,k} & = \frac{1}{N_{d,k}} \sum_{p \in P_{d,k}}x_{p}
 \ ,\label{eq:pixel-average}
\end{align}
 where $N_{d,k}$ is the number of pixels in $P_{d,k}$.

Substituting \eqref{eq:pixel-average} into \eqref{eq:direction-variance0} and
 simplifying, we get
\begin{align}
 \sigma_d^2 & = \frac{1}{N}\left[\sum_{p \in \mathrm{block}} x_p^2
 - \sum_{k = 0}^{N_d - 1} \frac{1}{N_{d,k}}\left(\sum_{p \in P_{d,k}}
 x_p\right)^2\right]\ .\label{eq:direction-variance1}
\end{align}
The first term is constant.
The optimal direction, $d_\mathrm{opt}$, is then just
\begin{align}
d_\mathrm{opt} & = \max_d s_d\ ,\label{eq:direction-variance2}
\end{align}
where
\begin{align}
 s_d & = \sum_{k = 0}^{N_d - 1} \frac{1}{N_{d,k}}\left(
 \sum_{p \in P_{d,k}} x_p\right)^2\ .\label{eq:direction-variance3}
\end{align}
The decoder still selects a direction for blocks with no strong directional
 features.
The goal is to allow stronger filtering without crossing directional edges.

We use a ``conditional replacement filter'' to remove noise without blurring
 sharp edges, like a median or bilateral filter, but simpler and easier to
 vectorize with SIMD:
\begin{align}
y(n) & = x(n) + \frac{1}{W}\sum_{k=-M}^{k=M}
 w_k\mathrm{thresh}\left(x(n + k) - x(n), T\right)
 \ ,\label{eq:conditional-replacement-diff}
\end{align}
 with the threshold function
\begin{align}
 \mathrm{thresh}(d, T) & = \left\{ \begin{array}{ll}
 d, & \lvert d \rvert < T\ ,\\
 0, & \mathrm{otherwise}.
\end{array}\right.\label{eq:threshold-function}
\end{align}
The result is a filter where the pixels whose difference from the center pixel,
 $x(n)$, exceed the threshold $T$ are simply replaced by the value of $x(n)$.
This keeps the normalization weight, $W$, constant, and setting it to a power
 of two avoids the division.

The first, or ``directional'' filter is the 7-tap conditional replacement
 filter
\begin{multline}
 y(i,j) = x(i,j) + \frac{1}{W}\sum_{k=1}^{3} w_{k}\Bigl[
 \mathrm{thresh}\left(x(i,j) -x\left(i + \left\lfloor kd_y \right\rfloor ,
 j + \left\lfloor kd_x \right\rfloor \right), T_d\right)\Bigr. \\
 \Bigl. {} + \mathrm{thresh}\left(x(i,j)
 - x\left(i - \left\lceil kd_y \right\rceil ,
 j - \left\lceil kd_x \right\rceil \right),
 T_d\right)\Bigr]\label{eq:directional_filter}
\end{multline}
 where $(i,j)$ is the pixel location, $d_x$ and $d_y$ are defined in
 Table~\ref{tab:Direction-parameters} and $T_{d}$ is the threshold for the
 directional filter stage.
Since the direction is constant over $8\times 8$ blocks, all operations in this
 filter are directly vectorizable.

We choose the weights $w_k$ to be
 $\mathbf{w} = \left[\begin{array}{ccc} 3 & 2 & 2\end{array}\right]$ with
 $W = 16$.
Although ringing is \textit{roughly} proportional to the quantization step
 size, $Q$, as the quantizer increases the error grows less than linearly
 because the unquantized coefficients become very small compared to $Q$.
We start with a power model of the form
\begin{align}
T_0 & = \alpha_{1}Q^{\beta}\ ,\label{eq:setting-T0}
\end{align}
 with $\beta=0.842$ and $\alpha_{1} = 1$, which were chosen by manually testing
 thresholds for $Q=5$ and $Q=400$ (on a linear scale).
We can use a stronger filter on more directional blocks, both because they have
 more ringing, and because there is less chance of blurring non-directional
 features.
Blocks that are less directional require a weaker filter.
We estimate the degree of directionality, $\delta$, as the difference between
the optimal variance and the variance along the orthogonal direction:
\begin{align}
\delta & = \lvert\sigma_{d_\mathrm{opt}}^2 - \sigma_{(d_\mathrm{opt} + 4)\,mod\,8}^2\rvert
 = s_{d_\mathrm{opt}} - s_{(d_\mathrm{opt} + 4)\,\mathrm{mod}\,8}\ .
\end{align}
The final threshold is then
\begin{align}
T_d & = T_0 \cdot \max \left(\frac{1}{2}, \min\left(3,
 \alpha_2\left( \delta \cdot \delta_\mathrm{sb} \right)^{0.16}
 \right) \right)\ ,\label{eq:setting-Td}
\end{align}
 where $\delta_\mathrm{sb}$ is the average of $\delta$ over a $32\times 32$
 superblock and $\alpha_2 = 1.02$.

The second filter stage is always horizontal or vertical, and operates across
 the directional lines used in the first filter:
\begin{multline}
 z(i,j) = y(i,j) + \frac{1}{W}\sum_{k=1}^{2} w_{k}\Bigl[
 \mathrm{thresh}\left(y(i,j) -y\left(i + \left\lfloor kd_y \right\rfloor ,
 j + \left\lfloor kd_x \right\rfloor \right), T_d\right)\Bigr. \\
 \Bigl. {} + \mathrm{thresh}\left(y(i,j)
 - y\left(i - \left\lceil kd_y \right\rceil ,
 j - \left\lceil kd_x \right\rceil \right),
 T_2(i,j)\right)\Bigr]\label{eq:directional_filter2}
\end{multline}
 where $d_x$ and $d_y$ are defined in Table~\ref{tab:Ortho-parameters} and
 $T_2(i,j)$ is a position-dependent threshold for the second stage.
Since the second filter risks blurring edges, and its input has less ringing
 than the first, it only has 5 taps and a conservatively chosen $T_2(i,j)$:
\begin{align}
T_2(i,j) & = \min\left(T_d, \frac{T_d}{3}
 + \lvert y(i,j) - x(i,j)\rvert\right)\ .\label{eq:setting-T2}
\end{align}
We choose the filter weights to be
 $\mathbf{w} = \left[\begin{array}{cc} 3 & 3\end{array}\right]$ with $W = 16$.

If a superblock was skipped and is not in an intra frame, it is never deringed.
Otherwise, a flag enables or disables deringing for a superblock.
Even when enabled, we do not dering $8\times 8$ blocks that were skipped and
 whose surrounding $4\times 4$ neighbors were also skipped (taken into account
 because of lapping).
The deringing process may read pixels that lie outside the current superblock.
When they lie another superblock, we use unfiltered pixel values---even for the
 second stage filter---to avoid adding a dependency between superblocks.
This allows filtering all superblocks in parallel.
When they lie outside the image, the threshold function,
 $\mathrm{thresh}(d, T)$, returns $0$.


\Section{Subjective Results}

\begin{figure}[t]
\begin{center}
\subfigure[Woman.]{
 \label{fig:pcs2015-woman}
 \includegraphics[width=2.4in]{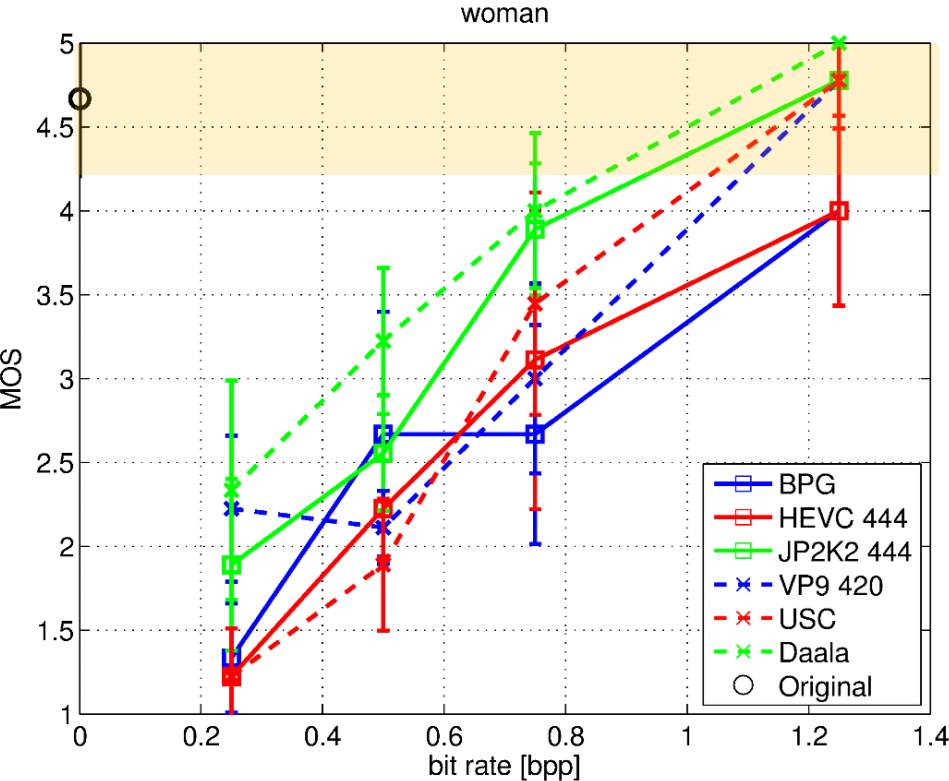}
}
\subfigure[Bike.]{
 \label{fig:pcs2015-bike}
 \includegraphics[width=2.4in]{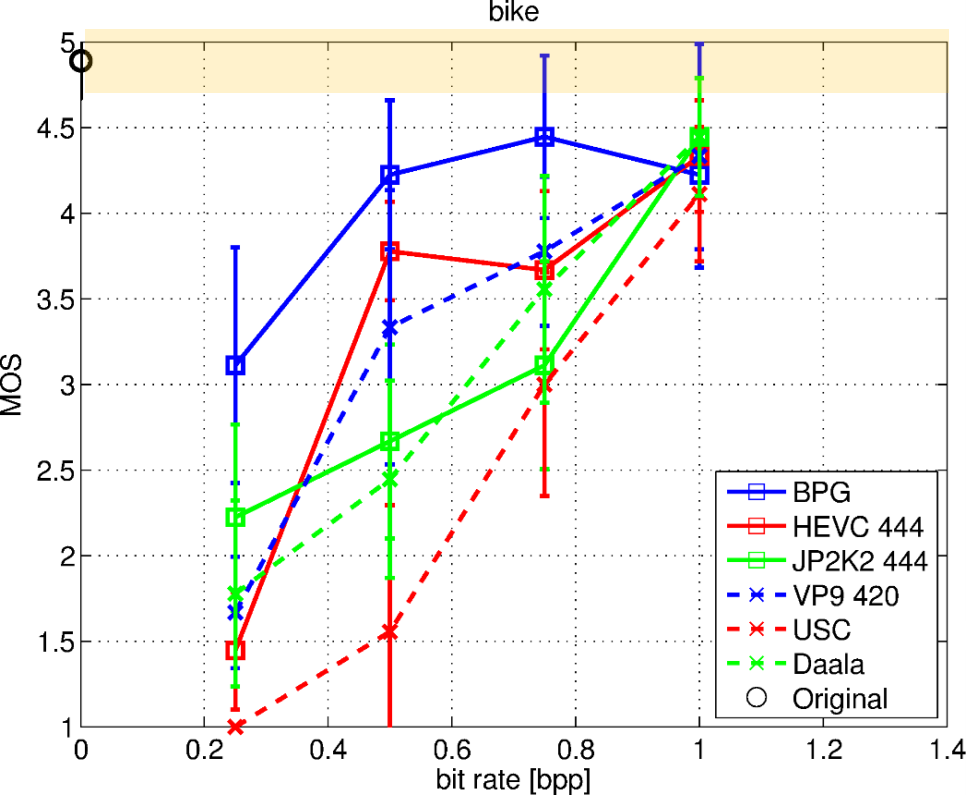}
}
\end{center}
\caption{\label{fig:pcs}%
Subjective still-image comparisons of Daala and several competing codec
 implementations.
Shown here are results from the images that gave the
 \ref{sub@fig:pcs2015-woman} best and \ref{sub@fig:pcs2015-bike} worst results
 for Daala.
}
\end{figure}

Daala participated in the 2015 Picture Coding Symposium evaluation of existing
 and future still image codecs~\cite{pcs2015}.
This included both objective evaluation with different metrics than those we
 used during development and subjective evaluation with test subjects drawn
 from a pool of multimedia quality experts.
Daala was compared with five other state of the art codecs including
 BPG\cite{bpg} and VP9\cite{vp9}.
Figure~\ref{fig:pcs} shows the subjective results for two of the six still
 images tested.

Our best results were on {\tt woman}, a close up portrait of a woman wearing a
 knit shirt.
About two thirds of this image are hair and skin, which contain texture that is
 well coded using PVQ even at very low rates.
Activity masking hides quantization errors in the hair where other codecs end
 up removing detail by zeroing high frequencies.

The image where Daala had the worst result was {\tt bike}, a picture of
 a tennis racket leaned against a bike wheel.
This picture contains strong directional edges which are well predicted by
 directional intra prediction.
Because Daala only contains a limited set of horizontal and vertical intra
 predictors, it spends more bits coding regions that are well predicted in
 other codecs.
The version of Daala tested by the competition was also using an earlier
 prototype of the deringing filter that was less effective.

\Section{Objective Results}

\begin{figure}[t]
\begin{center}
\subfigure[PSNR-HVS-M.]{
 \label{fig:graph-psnrhvs}
 \includegraphics[width=1.8in]{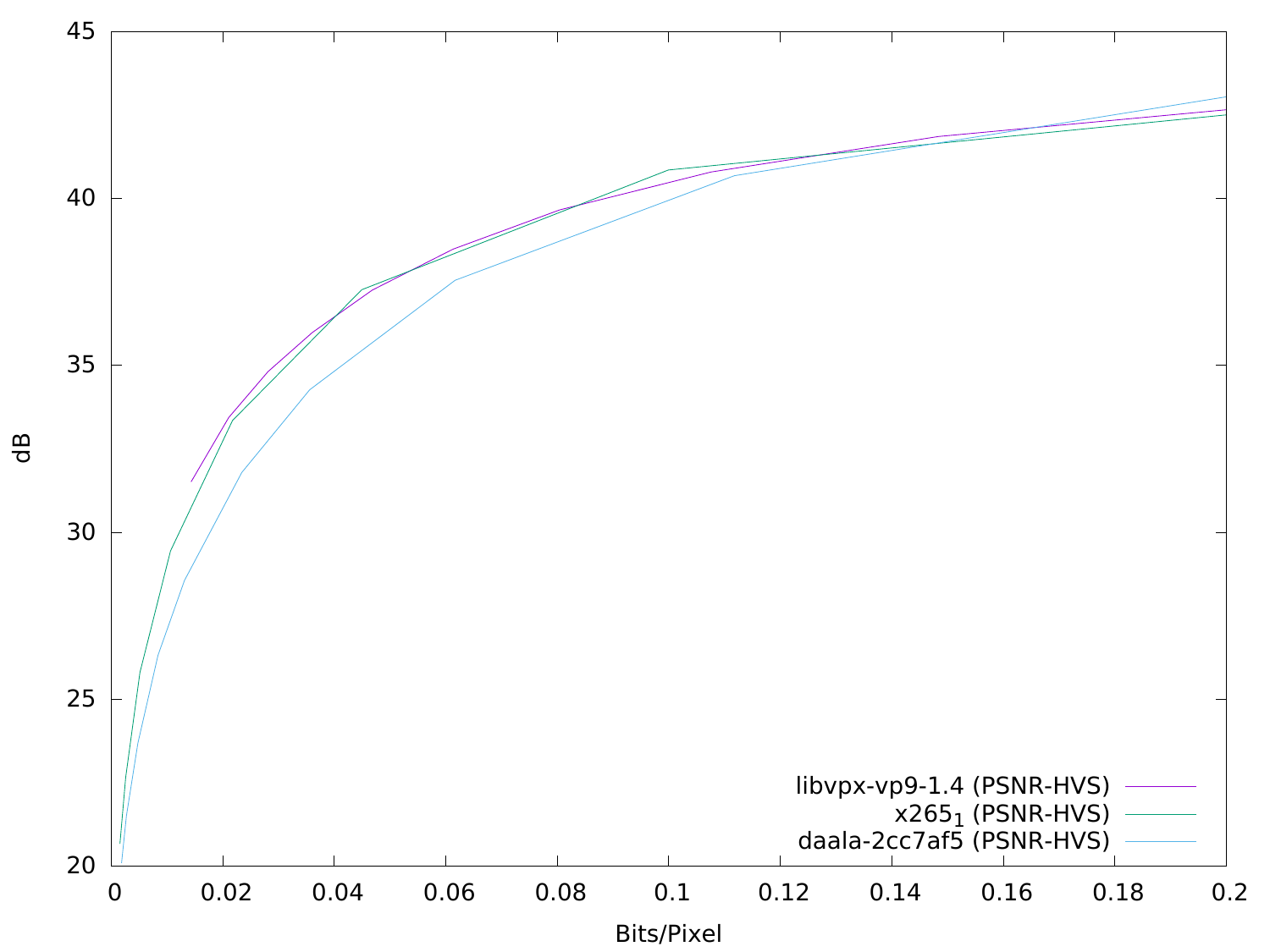}
}
\subfigure[Multiscale FastSSIM.]{
 \label{fig:graph-fastssim}
 \includegraphics[width=1.8in]{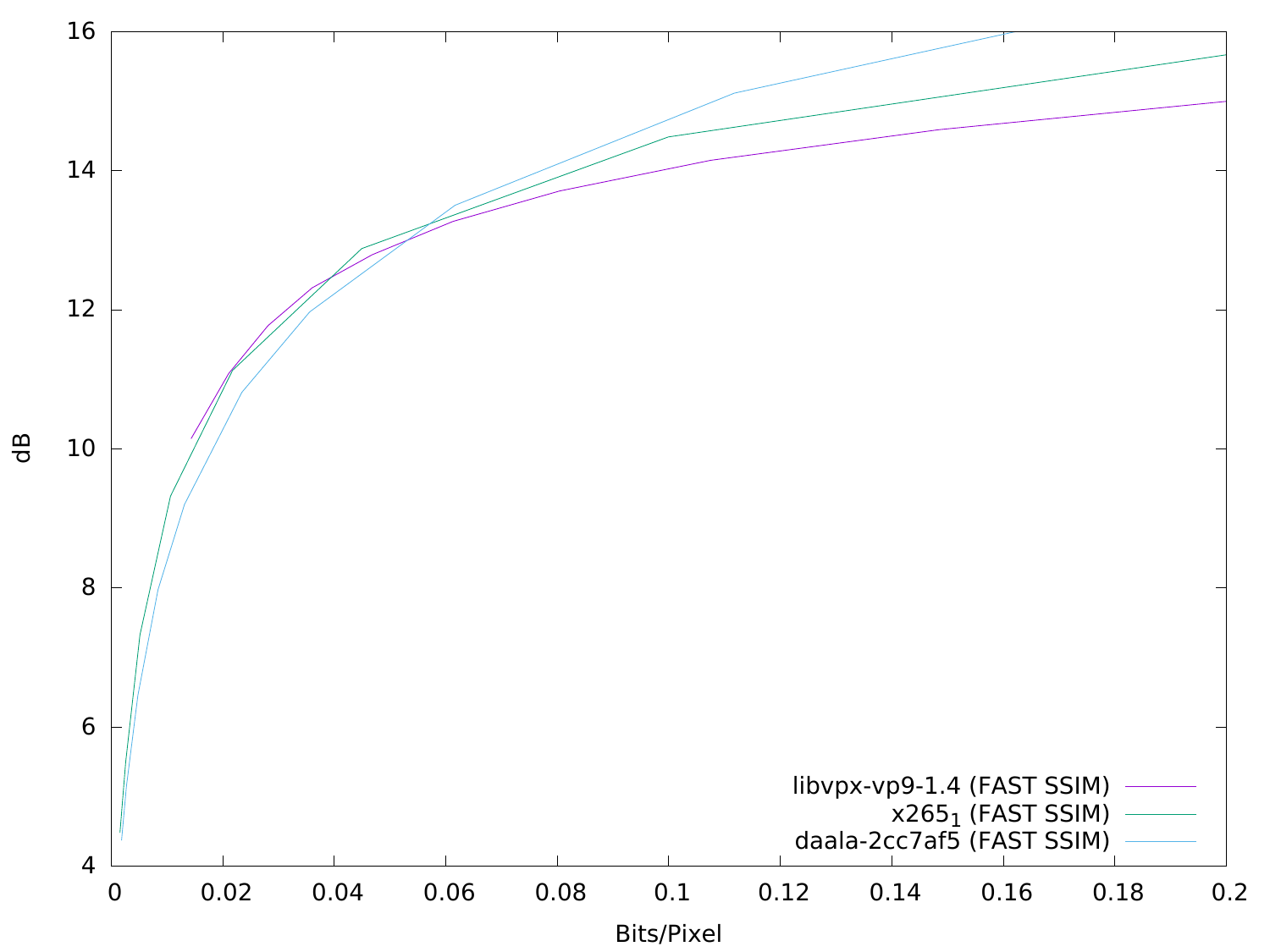}
}
\subfigure[PSNR.]{
 \label{fig:graph-psnr}
 \includegraphics[width=1.8in]{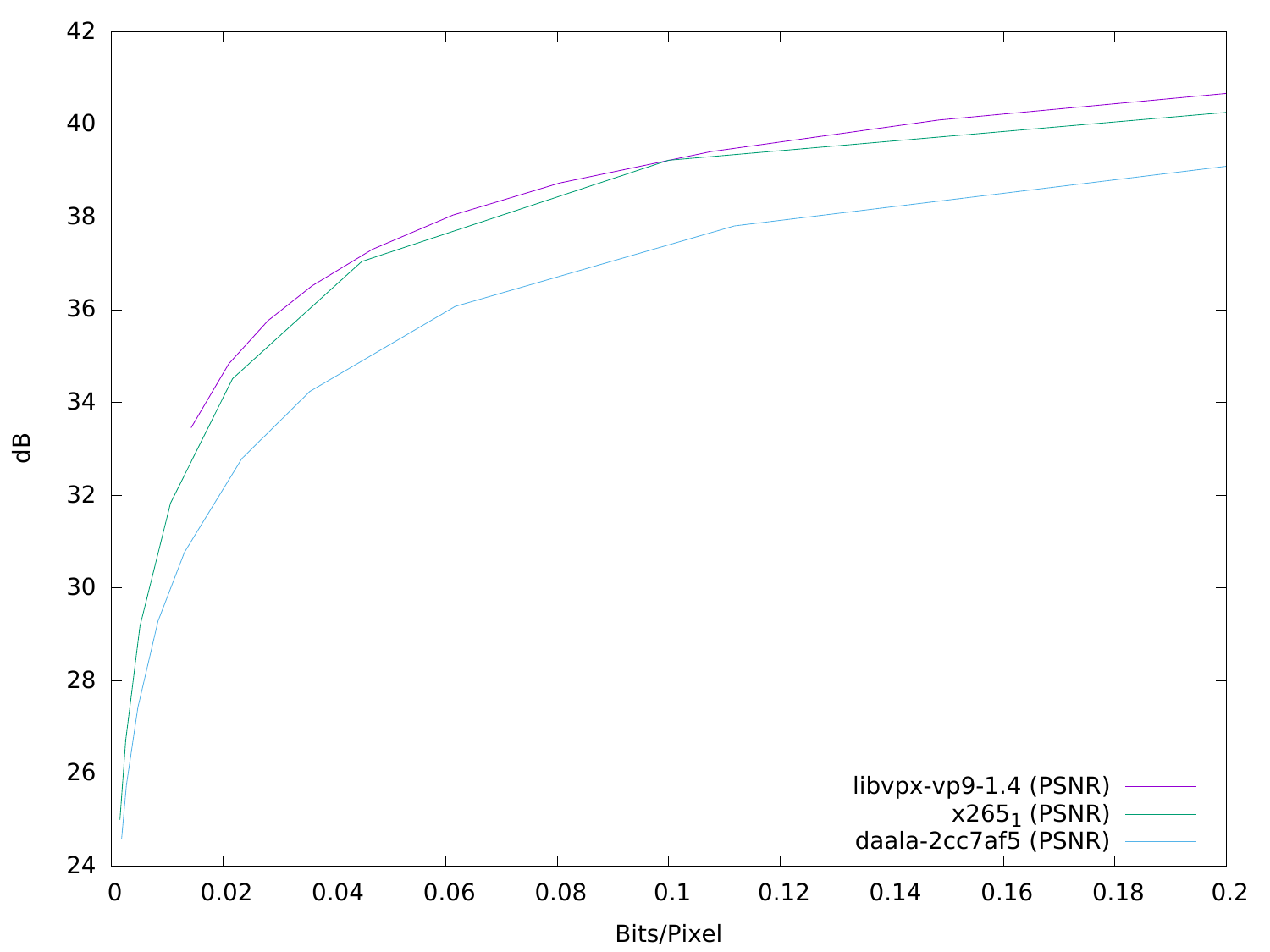}
}
\end{center}
\caption{\label{fig:object-metrics}%
Metric comparisons between Daala, libvpx-vp9 1.4.0, and x265 1.6.
}
\end{figure}

We computed objective results using standard metrics, using implementations in
 the Daala repository.
Daala is compared against x265 1.6 and libvpx-vp9 1.4.0.
The comparisons are made automatically by our open-source AreWeCompressedYet
 tool~\cite{AWCY}.
Two perceptual metrics are used, PSNR-HVS-M and multiscale
 FastSSIM~\cite{CB11,PSECAL07}, as well as PSNR.
Daala does much better than the other two codecs at high bitrates, though there
 is still room for improvement at lower rates.
Daala performs especially well on the perceptual metrics compared to PSNR, as
 expected when using perceptually-based coding methods.
As of this writing, we do not code out-of-order frames (B-frames or alt-refs),
 so we see substantial room for improvement on video.

\Section{References}
\bibliographystyle{IEEEtran}
\bibliography{daala-dcc16}

\end{document}